\documentclass[preprint2]{aastex}

\slugcomment{Accepted for publication in ApJ}

\newcommand{\ltsima} {$\; \buildrel < \over \sim \;$}
\newcommand{\gtsima} {$\; \buildrel > \over \sim \;$}
\newcommand{\lta} {\lower.5ex\hbox{\ltsima}}
\newcommand{\gta} {\lower.5ex\hbox{\gtsima}}

\shorttitle{The nuclei of LINERS}
\shortauthors{Chiaberge et al.}

\begin{document}

\title{The   HST  view  of   LINERs  nuclei:   evidence  for   a  dual
population?\thanks{Based   on  observations   obtained  at   the  Space
Telescope Science  Institute, which is operated by  the Association of
Universities  for  Research  in  Astronomy, Incorporated,  under  NASA
contract NAS 5-26555.}}


\author{Marco Chiaberge\altaffilmark{2}}
\affil{Istituto di Radioastronomia del  CNR - Via P. Gobetti
101, I-40129 Bologna, Italy}
\altaffiltext{2}{Space Telescope Science Institute, visiting researcher}
\email{chiab@ira.cnr.it}
\author{Alessandro Capetti}
\affil{Istituto Nazionale di Astrofisica (INAF) -- Osservatorio  
Astronomico di Torino, Strada Osservatorio  20, I-10025 Pino Torinese,
Italy}
\author{F.~Duccio Macchetto\altaffilmark{3}}
\affil{Space Telescope Science Institute, 3700 San Martin Drive,
Baltimore, MD 21218}


\altaffiltext{3}{Space Telescope Division, ESA}


\begin{abstract}

We study a  complete and distance-limited sample of  25  LINERs, 21 of
which have  been imaged with  the  Hubble Space Telescope.   In  nine
objects  we detect an   unresolved  nucleus.  In order to study
their physical properties, we compare the radio and optical properties
of the nuclei  of LINERs with those  of  other samples of local  AGNs,
namely Seyfert galaxies and low-luminosity radio galaxies (LLRG).  Our
results show that the  LINERs population is  not homogeneous, as there
are  two subclasses: i)  the  first class  is  similar to LLRG, as  it
extends the population of radio-loud nuclei to lower luminosities; ii)
the second   is similar  to Seyferts, and extends  the properties
of  radio-quiet nuclei towards   the lowest luminosities.  The objects
are optimally discriminated in the plane formed by the black hole mass
vs. nuclear radio-loudness:  all radio-loud LINERs have $M_{BH}/M\odot
\gta 10^8$,  while Seyferts and radio-quiet  LINERs have $M_{BH}/M\odot 
\lta 10^8$. The different nature of  the various classes  of local AGN are
best understood when the   fraction of the Eddington  luminosity  they
irradiate $L_o/L_{Edd}$  is plotted against the nuclear radio-loudness
parameter:   Seyferts  are   associated  with  relatively  {\it  high}
radiative efficiencies  $L_o/L_{Edd} \gta 10^{-4}$ (and high accretion
rates onto {\it low} mass black  holes); LLRG are associated with {\it
low} radiative  efficiencies (and low accretion  rates onto {\it high}
black hole  masses);  all LINERs  have  low radiative efficiency  (and
accretion rates), and  can be radio-loud  or radio  quiet depending on
their black hole mass.

\end{abstract}


\section{Introduction}


LINERs  are  particularly  important  as  they  are  the  most  common
manifestation  of nuclear activity  in nearby  galaxies, and  they may
represent  the  "missing  link"  between AGN  and  "normal"  galaxies.
Originally defined as  a class by Heckman (1980),  these galaxies have
optical spectra  dominated by  emission lines of  moderate intensities
and arising from gas in lower ionization states than in classical AGN.
Unambiguous     LINERs    have     [NII]6583/H$\alpha     >0.6$    and
[OIII]5007/H$\beta   <3$  (unlike   HII  regions)   or,  equivalently,
[OIII]/[OII]  $<1$ (unlike  Seyferts).  LINERs  are detected  in about
30\%  of statistically  complete  samples of  nearby objects  (Heckman
1980, Ho et al.  1997), but  despite two decades of studies the origin
of this activity is still unclear and there is no general consensus on
the physical processes at work in these galaxies (see e.g.  Filippenko
2003,  and references  therein).  In  particular, the  physics  of the
central engine in these sources is still unknown, as our understanding
of LINERs is  significantly hampered by the general  lack of detection
of any nuclear continuum emission  in key spectral regions such as the
optical.

Recent results have revealed  that the majority  of LINERs (as well as
Seyferts)  host  a compact radio  core with  a  flat  ($\alpha < 0.5$,
$\alpha$ is  defined as $F_\nu  \propto \nu^{-\alpha}$) radio spectrum
of high brightness  temperatures, the characteristic signature for the
existence of an AGN \citep{falcke,neil00,neil01,neil02}. If LINERs are
powered by  an active nucleus, the  accretion process in these objects
is likely to take place at an extremely low  level of activity.  Thus,
since  supermassive black holes   in galaxies  seem to be  ubiquitous,
LINERs  constitute a perfect laboratory for  studying accretion at its
lowest rates  and to  investigate the  connection  between galaxies in
which the central black  hole is ``active''  and those in which  it is
``quiescent''.

The presence  of non-thermal radio cores strengthens  the link between
LINERs  and  low-luminosity radio  galaxies  (LLRG),  which also  have
similar optical  spectra, and  opens the possibility  of investigating
the nature of  the former class following a  new approach, building on
the   results   we   obtained   for   LLRG.    In   \citet{pap1}   and
\citet{capettib2} we have  studied HST images of LLRG,  which show the
ubiquitous  presence of  unresolved optical  nuclei in  these sources.
The most striking  result of this study is that  the optical and radio
flux, as well  as the luminosity, correlate linearly  over four orders
of  magnitude and  with a  very  small scatter.   This led  us to  the
interpretation  that the optical  emission is  non-thermal synchrotron
radiation,  in analogy with  the origin  of the  radio core.   In this
scenario, the optical nuclear luminosity represents a firm upper limit
to any  thermal emission from  an accretion disk.  Adopting  a typical
black hole  mass of  $10^{9} M_\odot$ (e.g.   Macchetto 1999  and ref.
therein)  this translates  in a  fraction $<  10^{-4}-10^{-7}$  of the
Eddington luminosity.  The picture  which emerges is that accretion in
LLRG occurs  at a  very low accretion  rate and/or with  low radiative
efficiency,  as e.g.   in the  form of  advection  dominated accretion
flows  or similar  scenarios (ADAF,  ADIOS,  CDAF; e.g.   Rees et  al.
1982, Narayan \& Yi 1995, Narayan et al.  2000).  Thus, LLRG appear as
a substantially different manifestation  of the accretion process onto
a supermassive black hole than classical AGNs.

Some of  the key observational  questions that  will help to constrain
the physical framework for LINERs are the radio-optical properties and
the determination of the radiative efficiency of the accretion process
and/or the accretion  rate  around the  central  black  hole.  We  use
archival HST observations to address  these issues, by comparing their
properties with those  of the other local  unobscured AGN, i.e. Type~1
Seyferts and low luminosity radio galaxies.

In Section~2 we  describe the samples of  local  AGNs and  their black
hole mass    estimates; in section~3   we describe  the  HST images of
LINERs, and we discuss our method for the photometry of the nuclei; in
Section~4 we discuss our results and in Section~5 we draw conclusions.

Throughout this  paper we  adopt a  Hubble constant of  $H_0 =  75$ Km
s$^{-1}$ Mpc$^{-1}$ and $q_0 = 0.5$.

\section{LINERs in the framework of the local AGN population}

In order   to  investigate  the nature  of   LINERs  we compare  their
properties with those of other samples of  nearby AGN, namely Seyferts
and  low  luminosity radio  galaxies,   that,  when added  to  LINERs,
encompass  all  different manifestations of  nuclear  activity  in the
local  universe.  In the  following sections we  describe the selected
samples  and we  also gather  from  the literature  radio and  optical
nuclear fluxes as  well   as black  hole  mass estimates,   which  are
discussed in more detail below.

\begin{deluxetable}{l c c c c c r c}
\tablewidth{0pt}
\tablecaption{The sample of LINERS}
\tablehead{\colhead{Name}  & \colhead{Hubble Type} & \colhead{$S_{2cm}$} & \colhead{D} 
& \colhead{$\sigma$} & \colhead{Gal. $A_V$} &\colhead{BH mass} & \colhead{Ref.} \\
\colhead{~}  & \colhead{~} & \colhead{mJy} & \colhead{Mpc} 
& \colhead{Km s$^{-1}$} & \colhead{mag} &\colhead{M/M$_\odot$} & \colhead{~} }
\startdata
NGC~404     &    -3   &   $<$ 1.3 &  2.4 &  55 &  0.194 & $7.5\times 10^{5}$ &  0 \\
NGC~1052    &     -5  &    735$^*$    &    17.8  &   215 &  0.088 & $1.8\times 10^{8}$ &  0 \\
NGC~2681    &     0   &   $<$ 1.4 &    13.3  &   108 & 0.075 &  $1.1\times 10^{7}$ &  0 \\
NGC~2787    &     -1  &    11.4   &    13.0  &   206 &  0.434 & $3.9\times 10^{7}$     &  1  \\
NGC~2841    &      3  &     2.1   &    12.0  &   220 &  0.052 & $1.9\times 10^{8}$ &  0 \\
NGC~3368    &      2  &  $<$1.0   &     8.1  &   115 &  0.083 & $1.4\times 10^{7}$ &  0 \\
NGC~3718    &      1  &     10.8  &    17.0  &   178 &  0.047 & $8.4\times 10^{7}$ &  0 \\
NGC~4143    &     -2  &     10.0  &    17.0  &   243 &  0.042 &  $< 1.4\times 10^{8}$  &  2 \\
NGC~4111    &     -1  &  $<$ 1.1  &    17.0  &   157 &  0.048 & $5.1\times 10^{7}$ &  0 \\
NGC~4203    &     -3  &      9.0  &     9.7  &   152 &  0.040 & $< 2.3\times 10^{7}$  &  2 \\
NGC~4278    &     -5  &     87.7  &     9.7  &   251 &  0.095 & $3.4\times 10^{8}$ &  0 \\
NGC~4293    &      0  &      1.4  &    17.0  &   163 &  0.130 & $5.9\times 10^{7}$ &  0 \\
NGC~4314    &      1  &   $<$1.0  &     9.7  &   119 &  0.083 & $1.7\times 10^{7}$ &  0 \\
NGC~4394    &      3  &   $<$0.9  &    16.8  &   138 &  0.100 & $3.0\times 10^{7}$ &  0 \\
NGC~4438    &      0  &   $<$0.9  &    16.8  &   --  &  0.092 &    --     &  -- \\ 
NGC~4450    &      2  &      2.7  &    16.8  &   127 &  0.092 & $2.2\times 10^{7}$ &  0 \\
NGC~4457    &      0  &   $<$1.0  &    17.4  &   99  &  0.072 & $8.0\times 10^{6}$ &  0 \\
NGC~4548    &      3  &      1.6  &    16.8  &   135 &  0.126 & $<2.8\times 10^{7}$ &  2 \\
NGC~4550    &    -1.5 &    $<$0.7   &    16.8  &    91 &  0.129 & $5.7\times 10^{6}$  &  0 \\
NGC~4636    &     -5  &      1.8  &    17.0  &   209 &   0.092&  $1.6\times 10^{8}$ &  0 \\
NGC~4651    &      5  &  $<$ 1.1   &    16.8  &   --  &  0.089 & --        & -- \\
NGC~4736    &      2  &      1.7  &     4.3  &   126 &  0.059 & $2.1\times 10^{7}$ &  0 \\
NGC~4772    &      1  &      3.4  &    16.3  &   --  &  0.090 &    --     &  -- \\ 
NGC~4866    &     -1  &   $<$   1.1  &    16.0  &   210 &  0.091 & $1.6 \times 10^{8}$ & 0 \\
NGC~7217     &     2  &   $<$  0.6   &    16.0   &   127 &  0.292  & $2.2 \times 10^{7}$ & 0 \\

\enddata
\medskip

Radio fluxes are from \citet{neil02} except for NGC~1052
\citep{wrobel84} ($^*$ VLA observations at 1.4 GHz),  
NGC~404, NGC~4111 NGC~4651, NGC~4866 NGC~7217 \citep{neil00}; central velocity
dispersions are from  the LEDA database; Galactic  $A_V$ is from  NED.
References  for  BH masses   (column~8):  0 =   estimated  through the
correlation of \citet{tremaine}; 1 = \citet{ho02} 2 = \citet{sarzi02}.

\label{tabsample}

\end{deluxetable}

\begin{deluxetable}{l l c c r r r r}
\rotate
\tablewidth{0pt}
\tablecaption{HST observations of LINERs and nuclear data}
\tablehead{\colhead{Source Name}  & \colhead{Instrument/Filter} & \colhead{T$_{exp}$} & \colhead{Program ID} & \colhead{$F_\lambda$} & \colhead{$\log L_{7000\AA}$} & \colhead{$\log (L_r/L_o)$} & \colhead{$\log (L_o/L_{Edd})$} \\
\colhead{}  & \colhead{} & \colhead{s} & \colhead{} & \colhead{erg cm$^{-2}$
s$^{-1}$  \AA$^{-1}$} & \colhead{erg  s$^{-1}$
Hz$^{-1}$} & \colhead{} & \colhead{}}
\startdata
NGC~404     & WFPC2/F814W  & 320  & 5999 & $9.4 \times 10^{-16}$  &    24.96  &   $<$ 0.033 &     -4.398 \\
NGC~1052    & ACS/HRC/F330W & 300 & 9454 & $6.7 \times 10^{-17}$  &    25.25  &    4.152 &     -6.482 \\
NGC~2681    & WFPC2/F300W  &  1000 & 8632 & $2.6 \times 10^{-16}$ &     25.60 &  $<$ 0.884  &     -4.931 \\
NGC~2787    & WFPC2/F814W  & 730 & 6633 &  $<6.9 \times 10^{-17}$ & $<$25.39  &   $>$ 1.938 &    $<$ -6.349 \\
NGC~2841    &  --          & --  &   -- &  --       		  &     --    &   --     &       --   \\
NGC~3368    & ACS/HRC/F330W & 300 & 9454 &   $5.7 \times 10^{-17}$&     24.50 &   $<$ 1.356 &     -6.144   \\
NGC~3718    & NICMOS/F160W & 704 & 7330 &  $1.1 \times 10^{-16}$  &    26.11  &    1.419 &     -5.292 \\
NGC~4111    & WFPC2/F547M  & 300 & 5419 &  complex morphology     &     --    &    --    &       --   \\
NGC~4143    & WFPC2/F606W  & 400 & 8597 &  $1.5 \times 10^{-16}$  &    25.84  &    1.672 &   $>$  -5.783     \\
NGC~4203    & WFPC2/F814W  & 320  & 5999 & $2.3 \times 10^{-16}$  &    25.66  &    1.313 &    $>$ -5.178 \\
NGC~4278    & WFPC2/F814W  & 460 & 5454 &  $2.7 \times 10^{-17}$  &    24.75  &    3.232 &     -7.261 \\
NGC~4293    & WFPC2/F606W  & 160 & 5446 &  complex morphology     &      --   &     --   &      --    \\
NGC~4314    & WFPC2/F814W  & 600 & 6265 &  $<1.4 \times 10^{-17}$ & $<$ 24.46  &   $<>$ 1.574 &    $<$ -6.245 \\
NGC~4394    & WFPC2/F814W  & 320 & 9042 &  $<7.2 \times 10^{-17}$ & $<$ 25.65  &  $<>$ 0.816  &    $<$ -5.312 \\
NGC~4438    & WFPC2/F675W  & 1450 & 6791 &  complex morphology    &     --    &    --    &       --   \\ 
NGC~4450    & WFPC2/F814W  & 460 & 5375 &  $<1.1 \times 10^{-16}$ & $<$ 25.83  &     $>$1.11 &    $<$ -4.985 \\
NGC~4457    & --           & -- & -- &         --                 &     --    &     --   &        --  \\
NGC~4548    & --           & -- & -- &         --                 &     --    &      --  &       --   \\
NGC~4550    & WFPC2/F814W  & 1200 & 5375 & $<2.8 \times 10^{-16}$ & $<$ 25.25  &    $<>$1.118 &    $<$ -4.988 \\
NGC~4636    & WFPC2/F814W  & 200 & 8686 &  $<1.1 \times 10^{-18}$ & $<$ 23.84  &    $>$2.934 &    $<$ -7.844 \\
NGC~4651    & WFPC2/F814W  & 600 & 5375 &  $<3.2 \times 10^{-17}$ & $<$ 25.30 &    $<>$1.256 &       --   \\
NGC~4736    & ACS/HRC/F330W & 300  & 9454 & $2.2 \times 10^{-16}$ &    24.58  &    1.000 &     -6.228 \\
NGC~4772    & WFPC2/F606W  & 160 & 5446 &  $<2.6 \times 10^{-17}$ & $<$ 25.35  &    $>$1.965 &      --    \\ 
NGC~4866    &   --         & --  &  --  &  --                     &     --    &    --    &      --    \\
NGC~7217    & ACS/WFC/F814W & 120 & 9788 & $<1.3 \times 10^{-17}$ & $<$ 24.91  &    $<>$1.384 &    $<$ -5.909   \\
\enddata
\medskip

Fluxes  have been corrected for  extinction using Galactic
$A_V$ listed in Table~\ref{tabsample}. Objects that are plotted as double upper limits in the 
figures are marked with the symbol $<>$. 

\label{sampleobs}

\end{deluxetable}

\subsection{The selected samples of local AGN}

\subsubsection{LINERs}

Our LINERs sample contains 25 objects drawn from the Palomar survey of
nearby galaxies (Table\ref{tabsample}).   Their  catalog contains  486
bright ($B_T <12.5$) galaxies in the northern sky.  Here we consider a
complete, distance  limited  ($d<  19$ Mpc)  subsample  of  25 LINERs.
Transition sources  and  galaxies with   uncertain classification  are
excluded  from our  sample.   Radio   core  data  are  taken from   VLA
observations at 2cm \citep{neil00,neil02}.  For the prototypical LINER
NGC~1052 we have used observations  at 1.4 GHz from  \citet{wrobel84}.
We have assumed $\alpha = 0.0$ to convert the flux of NGC~1052 to 2cm,
in  agreement with the typical spectral  index of  compact radio cores
observed by \citet{neil01}.  $\alpha  =  0.0$ is  also assumed in  the
following sections to  convert the fluxes  to 5GHz in  order to derive
the nuclear radio  loudness parameter: a  change in the spectral index
of $\alpha=0.5$ would imply  a flux change   of less than a  factor 2,
leaving our results unaffected.   We have excluded  from our list  two
objects that  are part of  the Palomar catalog,  namely M~87 and M~84.
Although these galaxies are classified as LINERs by \citet{ho97}, they
are well known radio galaxies with extended radio  jets, and thus they
are included  in our  LLRG  sample.  Optical  observations of  the
central regions of 21/25  galaxies have been carried  out with HST and
the data are available in  the public archive.  Nuclear HST photometry
is described in the next Section.

\subsubsection{Seyferts}

Following \citet{hopeng},  we   consider a sample  of Seyfert~1  which
includes objects from  the Palomar survey \citep{ho97}, that
are faint and nearby ($D_{\rm med}= 20 Mpc$), together with relatively
brighter  (and  slightly  more distant)  objects   from the CfA survey
\cite{huchra83,ostermartel93}.    As noted     by   \cite{hopeng}, the
combined sample  itself  cannot  be  considered as   complete,  but it
provides a broader range   of nuclear luminosities, thus suitable   to
represent the Seyfert  phenomenon as a  whole.  We have  excluded from
the \citet{hopeng} sample those   classified as Seyfert 1.8  and  1.9,
since  the estimate  of  the nuclear  flux  in these  objects  may  be
affected by significant extinction  at optical wavelengths.  Clearly,
for   the same reason we   do not consider Type~2 Seyferts\footnote{We
have also removed NGC~1275  from the original  list, although this object
is often referred to  as  a Seyfert, since it   is a well known  radio
galaxy  (3C~84, Perseus~A)  with  a  peculiar emission  line spectrum.
However, this object is part of our LLRG sample.}.

HST/WFPC2 observations are  available for the  majority (25/32) of the
objects.  The data analysis and nuclear photometry has been performed
by \citet{hopeng}   by  modeling the  galaxy  brightness  profile with
multiple components using {\it GALFIT}. The radio  data are also taken
from the list in \citet{hopeng}.

\begin{deluxetable}{l r r c c r c}
\tablewidth{0pt}
\tablecaption{The sample of Seyfert~1}
\tablehead{\colhead{Name}  &  \colhead{$\log L_r$} &  \colhead{$\log L_o$} &\colhead{BH mass} & 
\colhead{Ref.}&  \colhead{$\log (L_r/L_o)$} &  \colhead{$\log (L_o/L_{Edd})$} \\
\colhead{~}  &  \colhead{erg s$^{-1}$
Hz$^{-1}$ } &  \colhead{erg s$^{-1}$
Hz$^{-1}$} &\colhead{M/M$_\odot$} & 
\colhead{~}&  \colhead{~}&  \colhead{~} }
\startdata
        IZw~1    &    29.28  &     29.36  &          --                &   -- &     0.028  &     -- \\
      MRK~205    &    29.12  &     29.23  &          --		       &  --  &    -0.070  &    -- \\  
      MRK~231    &    30.94  &     29.61  &          --		       &  --  &       1.338  &      --  \\
      MRK~279    &    29.02  &      --    &     $4.2 \times 10^{7}$    &   1  &      --      &      -- \\
      MRK~334    &    28.92  &      --    &          --		       &  --  &      --      &      --  \\
      MRK~335    &    28.54  &     28.67  &     $6.3 \times 10^{6}$    &   1  &    -0.090  &    -1.613\\
      MRK~471    &    28.31  & $>$ 27.58  &          --		       &  --  & $<$   0.751  &    --    \\
      MRK~530    &    29.21  &      27.9  &   $1.2 \times 10^{8}$      &   0  &       1.359  &    -3.659\\
      MRK~590    &    28.84  &     27.98  &    $1.8 \times 10^{7}$     &   1  &      0.903  &    -2.751\\
      MRK~744    &     27.9  &     --     &   $3.3 \times 10^{7}$      &   0  &     --       &    --     \\ 
      MRK~766    &     28.6  &     28.03  &          --		       &  --  &      0.581  &     --     \\ 
      MRK~817    &    28.98  &     28.54  &          --		       &  --  &      0.447  &     --      \\
      MRK~841    &    28.29  &     28.61  &          --		       &  --  &     -0.296  &     --      \\ 
      MRK~993    &    27.91  &     26.38  &          --		       &  --  &       1.634  &     --      \\ 
     MRK~1243    & $<$27.93  &      --    &          --		       &  --  &     --          &    --       \\ 
     NGC~3031    &    27.12  &     25.55  &     $6.3 \times 10^{7}$    &   1  &       1.714  &    -5.732 \\ 
     NGC~3227    &    28.17  &     27.26  &     $3.9 \times 10^{7}$    &   1  &      0.948  &    -3.812 \\ 
     NGC~3516    &    28.31  &     27.74  &     $2.3 \times 10^{7}$    &   1  &      0.642  &    -3.103 \\ 
     NGC~4051    &    27.5   &    26.84   &     $1.3 \times 10^{6}$    &   1  &      0.684  &    -2.754 \\ 
     NGC~4151    &    28.79  &     28.53  &    $1.5 \times 10^{7}$     &   1  &      0.316  &    -2.139 \\ 
     NGC~4235    &    27.89  &     26.81  &          --		       &   -- &       1.112  &     --    \\ 
     NGC~4395    &   24.97   &    24.33   &          -- 	       &   -- &      0.677  &     --    \\
     NGC~4639    &   26.26   &    25.25   &   $3.9 \times 10^{6}$      &   0  &       1.062  &    -4.829 \\ 
     NGC~5033    &   27.96   &    26.67   &   $2.8 \times 10^{7}$      &   0  &       1.316  &    -4.257 \\ 
     NGC~5273    &   26.95   &    26.26   &   $1.6 \times 10^{6}$      &   0  &      0.702  &    -3.415 \\ 
     NGC~5548    &    28.8   &    28.32   &    $1.2 \times 10^{8}$     &   1  &      0.501  &    -3.248 \\ 
     NGC~5940    &   28.17   &    28.11   &   $5.4 \times 10^{7}$      &   0  &      0.138  &    -3.102 \\ 
     NGC~6104    &$<$27.68   &    27.32   &   $4.0 \times 10^{7}$      &   0  &   $<$   0.383  &    -3.759 \\ 
     NGC~7469    &   29.39   &    28.49   &     $6.5 \times 10^{6}$    &   1  &      0.976  &    -1.804 \\ 
      UGC~524    &   28.51   &     --     &          -- 	       &   -- &      --      &     --    \\ 
     UGC~8621    &   27.75   &     --     &          -- 	       &   -- &      --      &     --    \\
    UGC~12138    &   28.57   &     --     &          -- 	       &   -- &      --      &     --    \\
\enddata
\medskip

Radio (column~2) and optical  nuclear  luminosity (column~3) are  from
\citet{hopeng}. Reference for  black  hole masses   (column~5): 1=from
\citet{ho02}, 0 = estimated through the correlation of
\citet{tremaine}.

\label{samplesy1}

\end{deluxetable}

\subsubsection{Low Luminosity radio galaxies}

To represent radio loud local  AGN, we consider the complete sample of
33 radio galaxies with  Fanaroff-Riley~I morphology (FR~I, Fanaroff \&
Riley  1974), which  is characteristic  of the  large majority  of low
luminosity  radio   galaxies  (LLRG),  taken  from   the  3CR  catalog
\citep{3cr}. The sample  is the same as in  \citet{pap1}.  The data we
use  in the  following are  summarized in  Table~\ref{samplefr1}.  The
objects span  a range of redshift $0.0037  < z < 0.29$,  with a median
value of $z=0.03$ which, for the cosmological parameters we are using,
corresponds  to 120  Mpc.  From the  point  of view  of their  optical
spectral properties,  FR~I have low intensity emission  lines and most
of them have LINER-like spectra.  Another  two samples of LLRG
(drawn  from the  B2 and  UGC  catalogs) with  slightly lower  average
redshift  have  been  studied with  the  HST \citep{capettib2,gijsugc}.
However, the 3C  sample best describes the properties  of the class of
radio galaxies  because it covers  a larger range in  radio luminosity
and it has the highest  detection rate of optical nuclei, an essential
ingredient   for  our   analysis.    Furthermore,  the   radio-optical
properties of  the B2  and UGC sample  are in complete  agreement with
those derived for the 3C.

\begin{figure}[h]
\epsscale{1} \plotone{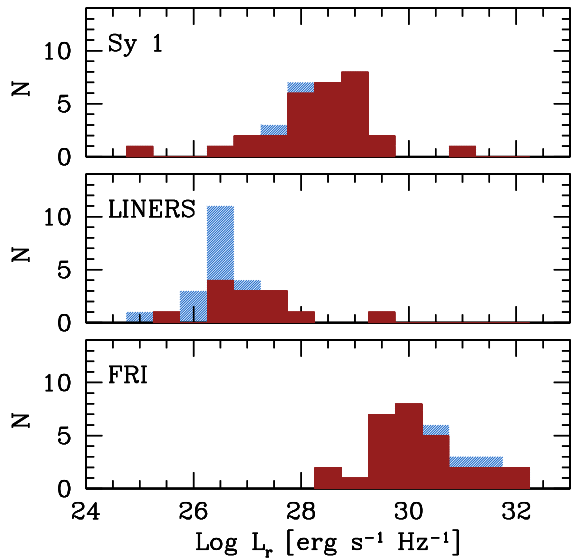}
\caption{Distances to the objects in the samples of Seyfert~1, LINERs 
and  FR~I radio galaxies.  Filled areas refer  to  objects in which an
optical core is detected.}
\label{dist}
\end{figure}

The  histograms of Fig.~\ref{dist} show the  distances  to the objects
belonging to the three selected samples.   Note that the three samples
have  been selected on the basis  of different properties, i.e.  radio
emission  for  the FR~I, optical  magnitude  of  the host and relative
intensity of  the emission lines for both  the LINERs  and Seyferts~1.
Therefore, the resulting   sample  cannot be considered   as complete.
However, the sources we consider  are representative of the properties
of AGN in the local universe.

\begin{deluxetable}{l c c c r r}
\tablewidth{0pt}
\tablecaption{Low Luminosity Radio galaxies from the 3CR catalog}
\tablehead{\colhead{Name} & \colhead{redshift} & \colhead{$\log$ BH mass} & 
\colhead{Ref.}&  \colhead{$\log (L_r/L_o)$} &  \colhead{$\log (L_o/L_{Edd})$}}

\startdata
         3C~28   &       0.19520 &    --   & - & $<>$2.583   &    --       \\
         3C~29   &       0.04479 & 8.121   & 0 & 3.992   &   -5.071\\
         3C~31   &       0.01690 & 8.581   & 0 & 3.575   &   -5.948\\
         3C~66B  &       0.02150  &    --   & - & 3.355   &    --   \\
         3C~75   &       0.02315 & 8.809   & 0 &  --     &    --   \\
        3C~76.1  &       0.03240 &    --   & - &   --    &    --\\
         3C~78   &       0.02900 & 8.660   & 0 & 3.395   &   -4.365\\
       3C~83.1   &       0.02510 &    --   & - & 3.963   &    --   \\
         3C~84   &       0.01700 & 8.134   & 0 & 4.238   &   -3.495\\
         3C~89   &       0.13860 & 8.785   & 0 & $>$5.176   &  $<$ -6.272\\
        3C~264   &       0.02060 & 8.568   & 0 &  2.980   &   -4.832\\
        3C~270   &       0.00737 & 8.716   & 1 & 4.516   &   -7.214\\
      3C~272.1   &       0.00370 &  9.079  & 1 &  3.207  &    -7.097\\
         3C~274  &       0.00370 &  9.531  & 1 &  3.734  &    -6.729\\
       3C~277.3  &       0.08570 &     --  & - &  3.698  &     --   \\
         3C~288  &       0.24600 &     --  & - &  4.419  &     --   \\
         3C~293  &       0.04520 &  7.994  & 0 &   --    &     --   \\
         3C~296  &       0.02370 &  8.742  & 0 &  4.142  &    -6.464\\
         3C~305  &       0.04100 &  7.927  & 0 &   --    &    --    \\
         3C~310  &       0.05400 &  8.041  & 0 &  4.146  &    -5.053\\
       3C~314.1  &       0.11970 &     --  & - &   2.810  &     --   \\
         3C~315  &       0.10830 &     --  & - &   --    &    --    \\
         3C~317  &       0.03424 &  8.498  & 0 &  4.333  &    -5.392\\
         3C~338  &       0.03030 &  8.849  & 0 &  3.809  &    -5.893\\
         3C~346  &       0.16200 &     --  & - &  3.768  &     --   \\
         3C~348  &       0.15400 &     --  & - &  3.884  &     --   \\
         3C~386$^*$ &    0.0170  &     --  & - &   --    &    --    \\
         3C~424  &       0.12700 &     --  & - &  3.867  &      --  \\
         3C~433  &       0.10200 &     --  & - &   --    &    --    \\
         3C~438  &       0.29000 &     --  & - &  4.416  &     --   \\
         3C~442  &       0.02620 &     --  & - &  3.129  &     --   \\
         3C~449  &       0.01810 &  8.547  & 0 &    3.100  &    -5.776\\
         3C~465  &       0.03010 &  9.082  & 0 &   3.940  &    -5.853\\
\enddata
\medskip

Redshifts  (column~2) are from  NED. Reference  for black  hole masses
(column~3): 1=from \citet{ho02}, 0 = estimated through the correlation
of \citet{tremaine}. 3C~28  has upper limits for both  the optical and
the radio  core, thus  we have  marked it with  the symbol  $<>$.  See
\citet{pap1} for radio cores and optical luminosities from HST images.
$^*$In 3C~386 a star is  superimposed to the nucleus, thus no optical
core flux can be measured \citep{pap4}.

\label{samplefr1}

\end{deluxetable}

\subsection{Black hole mass estimates}

\begin{figure}[h]
\epsscale{1} \plotone{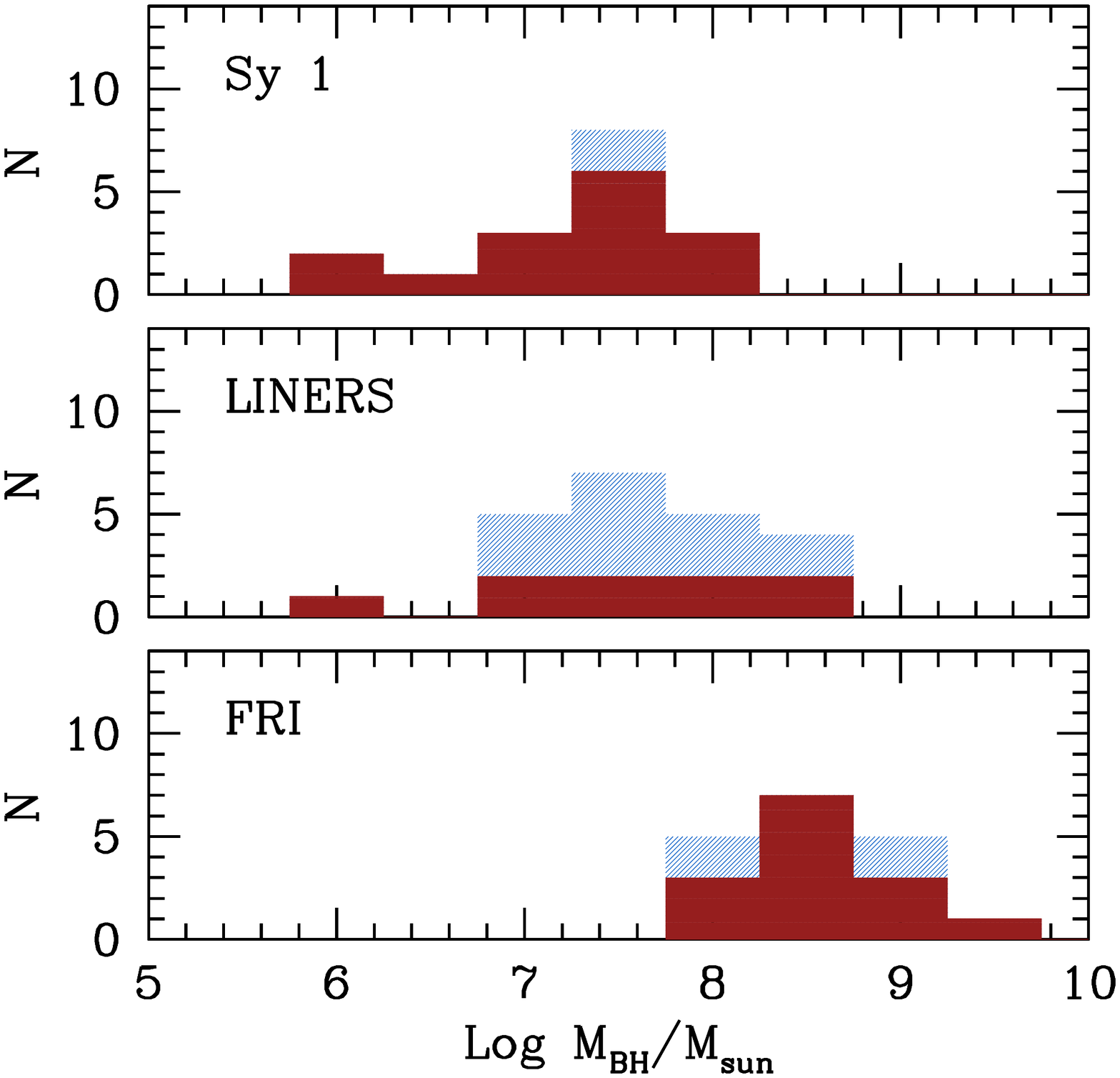}
\caption{Distribution  of black  hole  masses for  the three  samples.
Filled areas refer to objects in which an optical core is detected.}
\label{bh}
\end{figure}

In Fig.~\ref{bh} we show the distribution of black hole masses for the
three samples. When more   direct measurements are not  available, the
black  hole masses are estimated    through the correlation with   the
central  velocity  dispersion  determined  by  \citet{tremaine}.   The
velocity dispersions    have    been obtained from    the    HyperLEDA
database\footnote{e.g. http://leda.univ-lyon1.fr/}.  Of our samples of
local  AGN, twentytwo LINERs,  seventeen  Seyferts~1 and eighteen  FR~I
have black hole mass estimates (see Tables~\ref{tabsample} and
\ref{samplesy1}). 

LINERs appear to lie at intermediate values, between
Seyferts (low black hole mass, M$_{\rm BH} \lta 10^8$ M$_\sun$) and
radio  galaxies  (high  black  hole mass,   M$_{\rm BH}  \gta 10^8$
M$_\sun$) and all populations show a substantial overlap.

In the histograms of Fig.\ref{bh} we have  plotted as filled areas the
sources for  which  the optical  core is  detected,  to show that  the
presence of an optical  nuclear component is  not related to the black
hole mass.

\section{HST observations and LINERs nuclear fluxes}
\label{observations}

In this Section we derive measurements of the optical nuclear flux
for our sample of LINERs.
Twenty one out of twenty five  of 
the selected LINERs galaxies have been imaged with the HST as part of
several different programs and  therefore they have been observed with
different instruments, filters and  exposure times. The summary of the
observations for LINERs is in  Table\ref{sampleobs}, while we refer  to
the original papers for Seyferts and radio galaxies.

We perform optical photometry  of LINERs nuclei  using the same method
as for the FR~I sample, based on the analysis of the radial brightness
profile and described in detail in \citet{pap1} (hereafter CCC99).  We
also  checked for consistency against the  results obtained using {\it
GALFIT} \citep{penggalfit}. The fluxes derived   with the two  methods
are consistent to within 1 and 2 $\sigma$.

Of   the 21 LINERs imaged with   the HST, we  detect  the nucleus in 9
objects,  namely   NGC~404, NGC~1052,  NGC~2681,   NGC~3368, NGC~3718,
NGC~4143, NGC~4203, NGC~4378 and NGC~4736. For NGC~4111, NGC~4293, and
NGC~4438 the estimate of  the nuclear optical  emission is hampered by
the presence of lasrge scale dusty structures  along the line of sight
to the  nucleus.  

A few details on  the nuclear measurements  are worth mentioning.  For
NGC~3368 the nucleus is not detected in the  V band, but it is clearly
seen  in  the UV. We  argue  that this is  due  to the higher contrast
between  the  nucleus  and  the host  galaxy  stellar  emission  at UV
wavelengths compared with the V band.  The upper  limit in the optical
band ($F_\lambda <   9.1   \times 10^{-17}$ erg  s$^{-1}$    cm$^{-2}$
\AA$^{-1}$) is however  in substantial agreement  with the observed UV
nuclear flux (within less  than 1$\sigma$).  Conversely, NGC~3718  has
no detected nucleus in  the optical, while this is  seen in the IR 1.6
$\mu$m NICMOS image.  Here the absence of  the optical nucleus is most
likely due to the  presence of an extended  nuclear dust lane, clearly
seen in the optical images.

Since the  majority of the objects  have been observed  with the F702W
filter,  we have  converted the  measured flux  densities  to 7000\AA,
assuming  an  optical  spectral  index  $\alpha  =1$  ($F_\nu  \propto
\nu^{-\alpha}$).  Note  that if  we leave the  spectral index  free to
vary in the range  $\alpha  = 0 -  2$  this translates into a   typical
uncertainty on the 7000\AA~~  flux of $\lta  30\%$. Only for the worst
cases  of  NGC~1052 and  NGC~4736, which have  been  observed with the
F330W filter, the uncertainty would be still  less than a factor of 2.
Thus, this does not produce any significant effect in our analysis.

\section{The radio--optical properties of LINERs}

As  demonstrated  by our  analysis    of radio galaxies  samples,  the
comparison between  radio and optical  properties of  AGNs represents a
powerful tool to explore  the nature of their nuclei. Here we follow
the same approach   to investigate  the  properties of   LINERs nuclei,
setting a comparison with the other classes of local AGN. 

\begin{figure*}
\epsscale{2} \plottwo{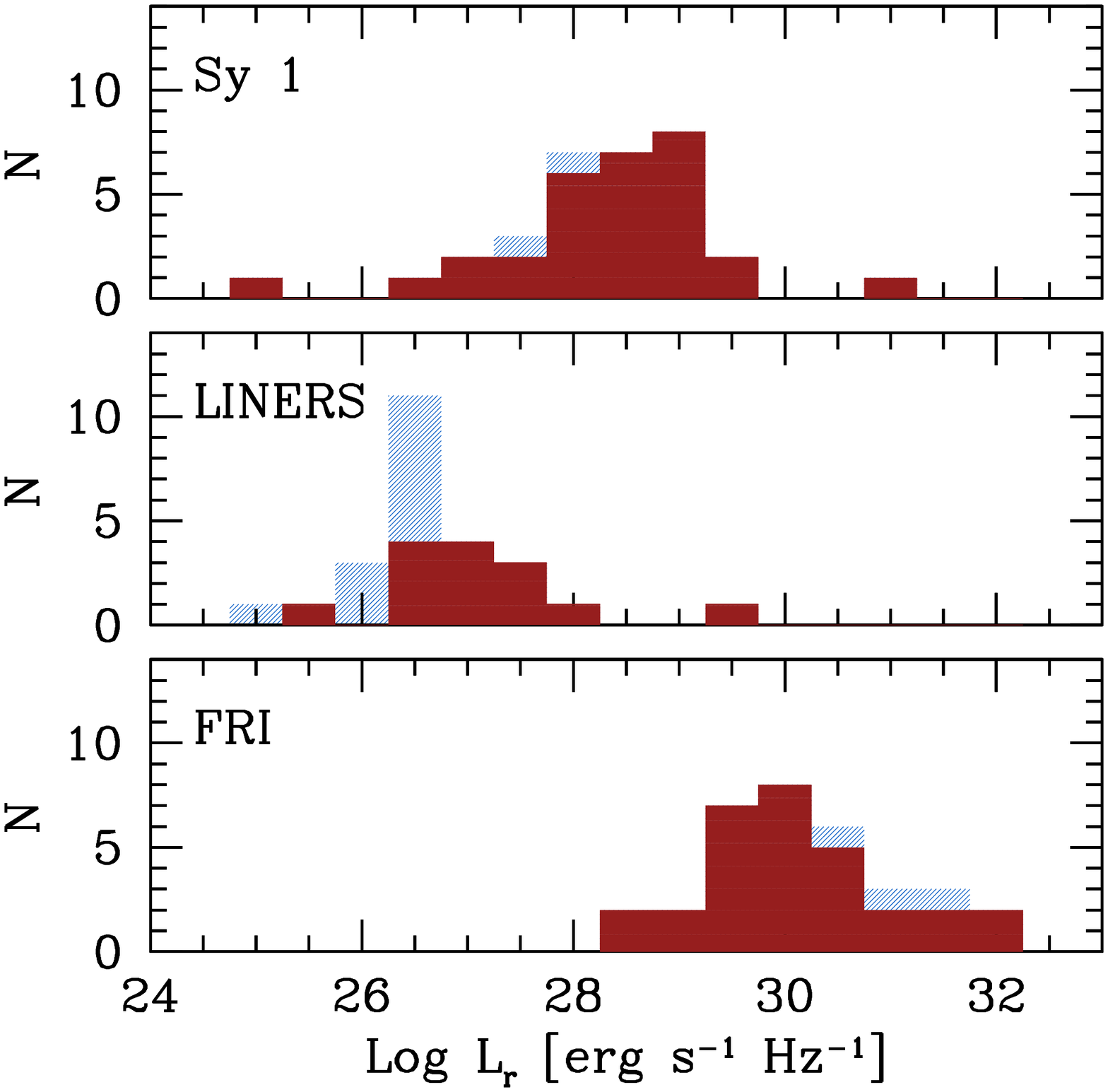}{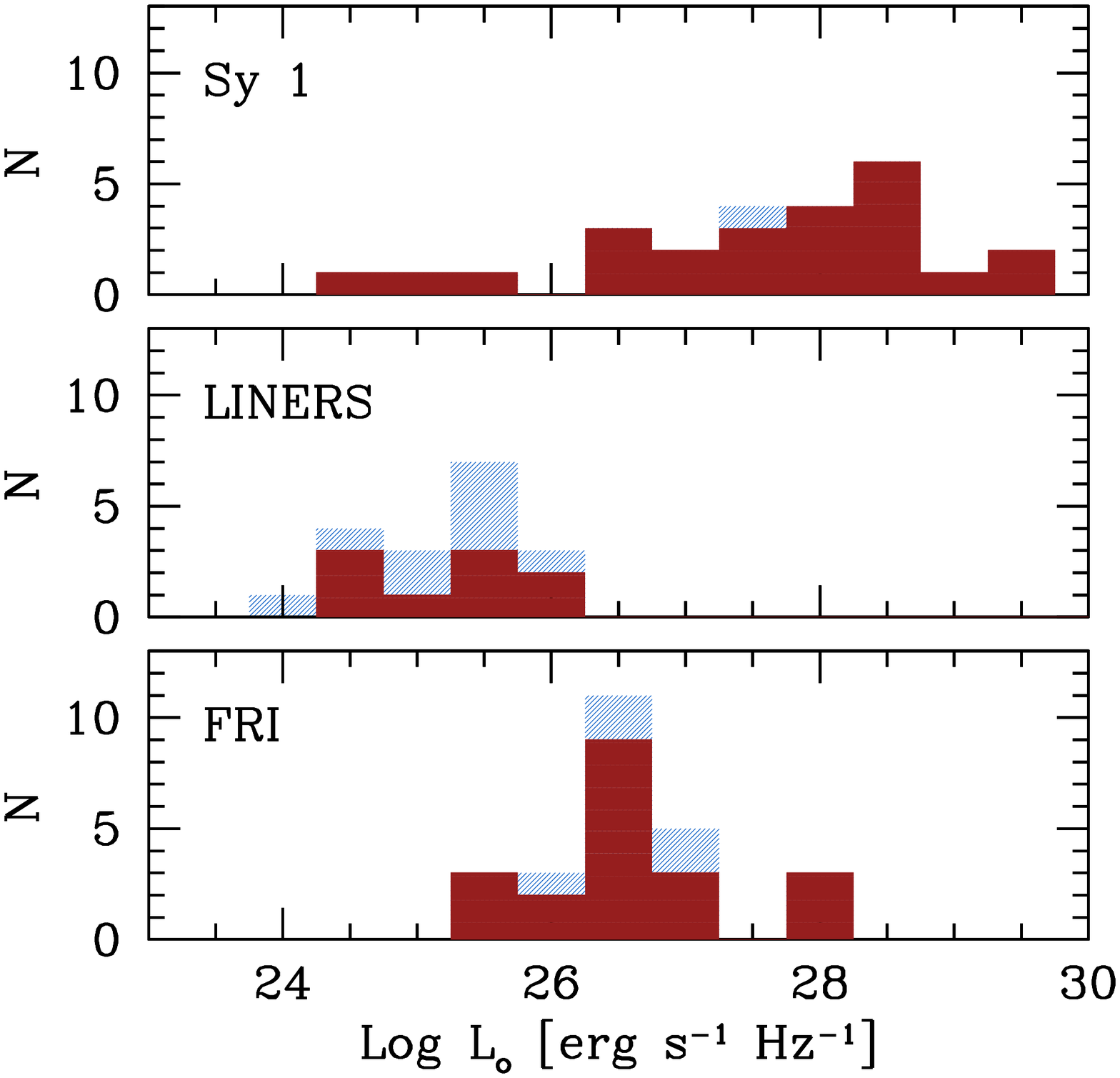}
\caption{Radio  core  (left  panel)  and optical  core  (right  panel)
luminosity for the LINER sample, compared with the Seyfert~1 nuclei of
\citet{hopeng}  and 3CR  FR~I. Shaded area refers to objects with 
an upper limit to the core emission.}
\label{lolr}
\end{figure*}

In Fig.  \ref{lolr}  we show the  histograms of the radio and  optical
core luminosity (left and  right panel,  respectively) for the  LINER,
Seyfert  and FR~I  samples.  LINERs' nuclei are substantially  fainter
than Seyferts and  FR~Is, both in optical and  radio luminosity.  More
specifically, the median   value  of the luminosity  distribution  for
LINERs' radio cores is a factor of $\sim 40$ and $\sim 3000$
fainter  than Seyferts and FR~I, respectively.   In the optical,
these  differences are  slightly  smaller and  there  is more  overlap
between the different classes: LINERs are on average a factor of $\sim
300$ fainter  than Seyferts, and only  a factor $\sim  6$ fainter than
FR~I.

\begin{figure}
\epsscale{1} \plotone{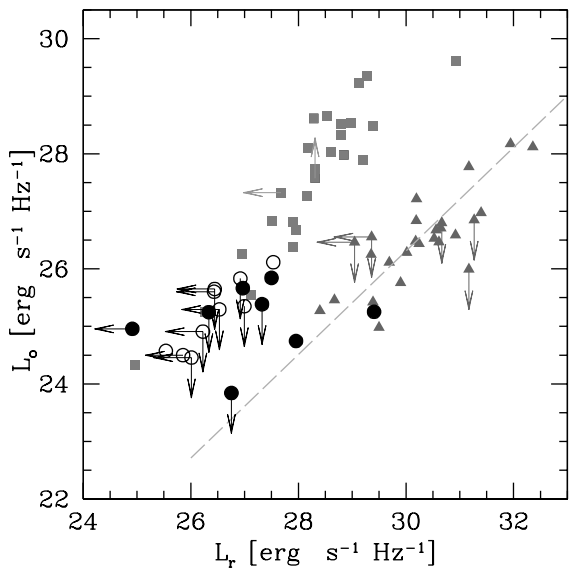}
\caption{Optical nuclear luminosity vs. radio core luminosity for 
LINERs   (circles), Seyferts    (squares)  and  FR~I  radio   galaxies
(triangles).  The   dashed  line is the    correlation between the two
quantities found  for 3CR  FR~I  sample. Open  circles  are LINERs  in
late-type hosts, filled circles are LINERs in early-type hosts.}
\label{lum}
\end{figure}

In Fig.  \ref{lum} we plot the  optical nuclear luminosity against the
radio core luminosity  for the three samples.  As  shown in CCC99, the
nature of optical and radio cores of FR~I nuclei is explained in terms
of  synchrotron emission  from the  base  of the relativistic jet, and
this   results  in  a tight   linear    correlation between these  two
quantities.  The Seyfert~1  nuclei lie in   a different region of  the
plane: for a similar radio luminosity, they show a significant optical
excess ($> 2$ dex).     While  a non-thermal synchrotron origin    for
compact radio  emission   in Seyferts has  been   clearly established
\citep{ulvestad,mundell,neil00}, the  optical  excess is  most  likely
thermal. In fact, this behavior is  qualitatively similar to that observed in
broad line radio galaxies, in which the  optical excess is interpreted
as a result of thermal emission from a radiatively efficient accretion
disk \citep{pap4,papuv}.

The nature of the LINERs'  radio cores is considerably more uncertain,
compared with e.g.  those of radio-galaxies.  However, several authors
\citep{falcke,neil00,neil01}  have recently  shown that,  similarly to
Seyferts,  most of  the LINERs'  radio emission  is in  the form  of a
compact, flat spectrum radio  core, the characteristic signature of an
AGN.  As  pointed out in  Sect.  \ref{observations} nine  galaxies in
our sample  also show  a central optical  component.  The  location of
these objects in the radio-optical plane is less homogeneous than both
LLRG and Seyferts.   While two of them (NGC~1052  and NGC~4278) lie on
the  correlation of FR~I  galaxies, the  remaining seven  objects with
detected  optical cores  show significant  optical excess,  similar to
Seyferts.  Those objects for which only upper limits are available are
located along a  general extension of the Seyfert  locus towards lower
luminosities, except for  NGC~4636 which lies a factor  of 3 above the
correlation, at its very low end and for which a strong optical excess
can be excluded.   In Fig.  \ref{lum} we have  also marked LINERs with
different host galaxy type with different symbols.  Filled circles are
representative of ellipticals, while empty circles represent late type
galaxies  (Hubble type  $\leq 0$  or $>0$,  respectively).   Among the
detected nuclei, it appears that the  two objects that lie on the FR~I
correlation have  elliptical hosts, while either host  galaxy type are
present in objects with optical excess.

The  location of  LINERs nuclei  in the  radio-optical plane  might be
affected  by obscuration in  the optical  band.  This  is particularly
worrisome for the objects that lie on the FR~I correlation, since they
might move towards the Seyfert's locus.  However, of those three LINERs
nuclei, NGC~4636 is undetected in the X-rays \citep{jones} as it is in
the  optical,  NGC~4278  has  $N_H< 0.035  \times  10^{22}$  cm$^{-2}$
\citep{terashima03}, while only  NGC~1052 shows significant absorption
($N_H = 9\times 10^{22}$  cm$^{-2}$, partial covering model, Terashima
et al.  2002).  However, its  nucleus is clearly detected  at 3300\AA.
We have  also checked that there  is no strong infrared  excess in the
HST/NICMOS image at $1.6\mu$, implying  that extiction does not play a
significant role  in the  IR-to-optical band.  The  resulting spectral
index is  $\alpha_{1.6-0.3\mu} \sim 2$, in agreement  with the typical
values of FR~I radio galaxies \citep{papuv}. As for the FR~I and other
classes of X-ray obscured AGN, a low optical extinction coupled with a
relatively  high  X-ray column  density  might  be  explained by  e.g.
unusual dust/gas  ratios \citep{granato} or  by the properties  of the
dust grains \citep{maiolino01}.

\subsection{The nuclear radio-loudness parameter}

\begin{figure}[h]
\epsscale{1} \plotone{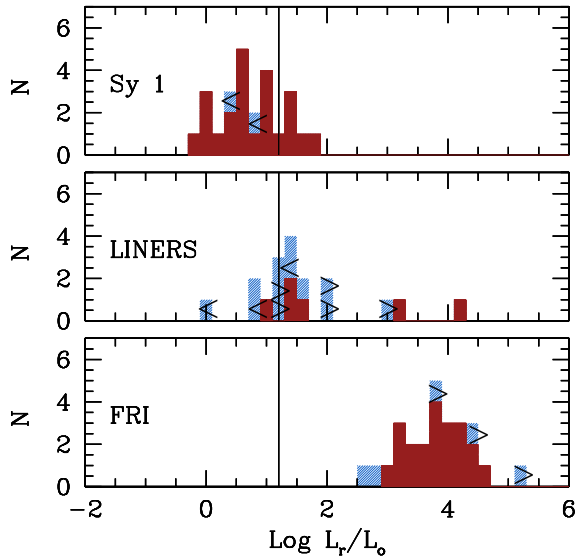}
\caption{Radio core to optical core flux ratio. Shaded area refer to object
with  upper limits  to the  nuclear  emission upper.  The solid  lines
correspond to the ``historical''  dividing line between radio-loud and
radio  quiet AGN.  Note that because  of the  different observing band
(R-band  instead  of  B-band)  our definition  of  that dividing  line
corresponds to $R \sim 16$.}
\label{rloud}
\end{figure}

A slightly different  path can be followed  by studying the connection
between radio  and   optical    properties  considering   a    nuclear
``radio-loudness'' parameter, defined as  $R= F_{5GHz}/F_R$,  i.e. the
ratio between radio and optical nuclear fluxes.  Although the presence
of a  clear dichotomy in the  radio-loudness distribution  of powerful
AGN is still a matter of  debate (e.g.  White  et al.  2000, Cirasuolo
et al.  2003),   Seyfert~1 and LINERs  are historically  considered as
radio-quiet with respect to their global properties.  Intriguingly,
\citet{hopeng}  have shown that when the  nuclear  emission of Seyfert
(type 1  through  1.9) is disentangled  from the  host  galaxy stellar
component, the  majority of their nuclei  have $R= F_{5GHz}/F_B > 10$,
i.e.    they fall into the standard   definition of radio-loud objects
\citep{kellermann89}.  But how  does this compare  with  the nuclei of
standard  radio-loud  objects?  In   Fig.   \ref{rloud}  we  show  the
histograms  of the radio-loudness  parameter  for Seyferts, LINERs and
FR~I
\footnote{Note that  with our definition of $R$  as $R= F_{5GHz}/F_R$,
our   results  are   slightly   different  from   those  obtained   by
\citet{hopeng}.   However, since  most of  the observations  have been
performed with optical R-band filers  (mainly F702W), we prefer to use
a  reference   wavelength  as  close  as  possible   to  the  original
observation, because of the uncertainty on the spectral indices of the
nuclei.  Therefore, for  the sake of coherence, in  our case the usual
definition of  radio-loud object would translate  into $F_{5GHz}/F_R >
15.8$ (assuming  an optical spectral  index $\alpha=1$).  None  of our
conclusions  are  affected by  this  choice  of  a slightly  different
spectral  band.}.  The  radio-optical  correlation found  for FR~I  is
linear  and  it  translates  into  a  ``radio-loudness''  distribution
clustered  around  $\sim 10^4$.   The  usual  value of  $F_{5GHz}/F_R$
adopted in  the literature as  the separation between  radio-quiet and
radio-loud object  does not  correspond to any  significant distiction
between Seyferts  and radio galaxies nuclei. Although  some Seyfert 1s
are indeed above this {\it threshold} the median value for Seyferts is
more than  a factor of $\sim  1000$ lower than that  measured in FR~I,
and thus they appear to be fundamentally different from LLRG.

The LINERs  span a  large range in  $R$. Some  of them are  similar to
Seyferts, while at  least three of the objects of  the sample have $R$
in the region spanned by bona fide radio loud objects.  Thus, when the
radio--loudness parameter is considered,  LINERs nuclei appear to link
the other two populations of local AGN. 
  
\begin{figure}[h]
\epsscale{1} \plotone{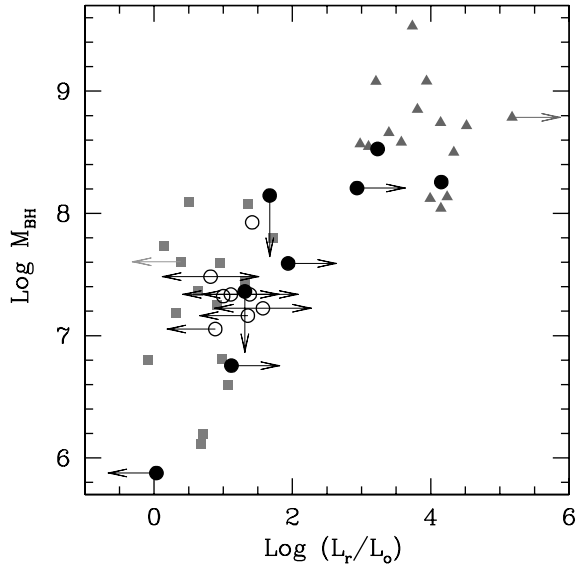}
\caption{Black hole mass estimate vs. radio loudness of the nuclei for
the three samples of local AGN. LINERs are marked as circles, Seyferts 
as  squares  and  FR~I  radio   galaxies
as triangles. Open  circles  are LINERs  in
late-type hosts, filled circles are LINERs in early-type hosts. }
\label{bhvsrl}
\end{figure}

The analysis of the different behavior of radio-loudness benefits from
considering also the black hole mass.  In Fig.\ref{bhvsrl} we plot the
black hole  mass versus the nuclear  radio loudness.  We  stress that,
since  the three samples  have been  selected  according to completely
different criteria and  observational  properties, it is  dangerous to
carry  out any  statistical test on   the total sample   of local AGN.
Thus,    no   correlation  can    be     derived between  these    two
quantities. Nonetheless, ``radio-loud'' nuclei appear in galaxies with
more massive black holes and separate themselves from the radio-quiet,
low black hole mass  objects.  Taken at face value,  from this plot it
appears that the  nuclei of local AGN  are ``radio-loud'' only if  the
mass of the black hole is $M_{BH} > 10^8  M_{\odot}$.  This is similar
to the results obtained by \cite{dunlop}  on the distribution of black
hole masses in  QSO.  Their analysis, which is  based on the QSO  host
galaxy luminosity, shows that radio-loud  objects are confined to  the
higher end of the black  hole mass distribution for  their sample.  On
the  other hand,  \citet{woourry} have  recently found  that the black
hole mass does not drive  the radio-loudness of powerful high-redshift
quasars:  in their large  sample radio  quiet  quasars with high black
hole masses do exist. However, the picture may change substantially in
the nearby universe, as shown by our results.

The  separation among  LINERs with  detected optical  nuclei  based on
``radio-loudness''  parameter alone is  confirmed and  strengthened by
our  analysis: the  three  LINERs with  large  value of  $R$ are  also
associated to the largest black holes.  Note however that the position
of  several LINER's nuclei  is partially  undetermined in  this plane,
since eight  of them are  not detected either  in the radio or  in the
optical, and for three of them  neither the optical nor the radio core
is detected.  However,  at least ten LINERs are  well constrained in
this plane, and for all  of these sources, their ``radio-loudness'' is
clearly related to their black hole mass.

This analysis suggests that there  may be two distinct populations  of
LINERs.   A first  group   (namely NGC~1052,  NGC~4278 and   NGC~4636)
extends the behavior of FR~I radiogalaxies to even lower powers.  Note
that they all show extended radio emission: NGC  1052 is associated to
a compact (2.8 kpc in size) double-lobed radio-source \cite{wrobel84};
NGC~4278 also shows extended radio emission \cite{wrobelheeschen}, but
on much smaller scales ($<100$ pc); NGC~4636 shows an extended ($\sim$
2kpc)   jet-like  structure    \citep{stanger}.    Their  total  radio
luminosities $\nu L_\nu$ at 1.4 GHz are  in the range $8\times 10^{37}
- 5\times 10^{38}$ erg s$^{-1}$.  As a reference, the radio luminosity
of  LLRG from  the 3C   spans the  range   $10^{38}$ to $10^{42}$  erg
s$^{-1}$.  In the following, we will refer to this sub-class of LINERs
as to ``radio-loud'' LINERs.

The second group of LINERs  (which possibly represents the majority of
the objects of the complete sample) behave like Seyfert~1 at the lower
end of  their luminosity function.  We  will thus refer to  them as to
``radio-quiet''  LINERs.   However,  the  number  statistics  are  not
sufficient to  test whether there  is a continuous  transition between
the two subclasses, or the population is bi-modal.

Note that   the    radio-quiet LINERs  are  associated  with   a mixed
population of host  galaxy morphologies,  while the three   radio-loud
LINERs are hosted by early type galaxies.  Interestingly, this is also
the case for LLRG, which are exclusively hosted by E or S0 galaxies.

The separation between the two classes of LINERs does not appear to be
related  to their optical  spectral  features, since faint, relatively
broad (FWHM $\sim  2000$ km s$^{-1}$)  emission lines (H$\alpha$) have
been detected  in the optical spectra of  several objects belonging to
either flavor of LINERS \citep{barth,hobroad}.

\subsection{On the properties of the accretion process}

From the estimates of the central black hole masses,  we can infer the
fraction  of   the Eddington  luminosity  at   which  the  nucleus  is
irradiating.   This quantity can then be  related to the properties of
the  accretion process  in the  different classes.   This estimate  is
uncertain for two main reasons: i) when using the correlation with the
central velocity  dispersion,  the   black  hole  mass  estimates  are
uncertain  by a  factor of $\sim   3$; ii) even  more importantly, the
optical   luminosity represents  only a    fraction of the  bolometric
luminosity of the   nucleus.  The  bolometric  correction derived  for
quasars \citep{elvis} ($L_{bol}/L_R \sim 15$) is probably sufficiently
accurate in the   case of Seyfert galaxies,  as  they show  a spectral
energy distribution quite similar to those  of their higher luminosity
counterparts \citep{alonso}.  Instead, as  already pointed  out above,
the  nuclear   emission of FR~I  is   best interpreted  as non-thermal
synchrotron radiation from the base of the jet (CCC99).  Therefore, in
that case,  $L_o$  represents an {\sl upper   limit} to  the radiation
produced by  the accretion  process.   For LINERs the picture  is even
more uncertain, since the nature of the spectral  shape of the nuclear
emission is poorly determined.  Bearing  in mind these considerations,
we do not attempt to perform any bolometric correction to the data and
we  preferred to   simply compare   optical nuclear luminosities    as
fraction of the Eddington luminosity.  Note that,  as it will be clear
in  the following, our conclusions  are not affected by an uncertainty
$\gta 1$ dex on the bolometric correction.

\begin{figure}[h]
\epsscale{1} \plotone{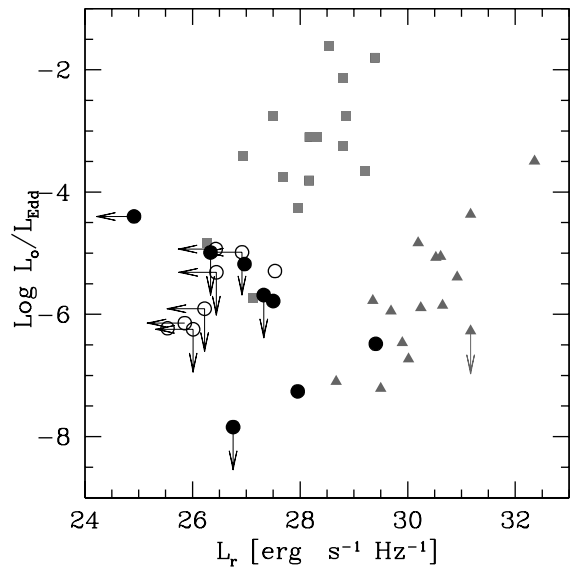}
\caption{$L_o/L_{Edd}$ vs. radio core luminosity of the nuclei for the 
three samples of local AGN. LINERs are marked as circles, Seyferts as  squares  and  FR~I  radio   galaxies
as triangles. Open  circles  are LINERs  in
late-type hosts, filled circles are LINERs in early-type hosts. }
\label{leddlr}
\end{figure}

In  Fig.\ref{leddlr} we plot  $L_o/L_{Edd}$  against $L_r$.   Compared
with the  diagram  of Fig.  \ref{lum}, the   sources are  more clearly
separated,  because  of  the different   average  black hole   mass of
different classes of  objects.  Seyfert nuclei  irradiate at a  larger
fraction of $L_{Edd}$ with respect to both LINERs and FR~I. 
For a  standard bolometric correction factor  of 15, the average value
for Seyfert's  $L_{bol}/L_{Edd}$  is in   the range  expected from   a
standard   optically thick accretion disk.   Only  two of them, namely
NGC~4639   and NGC~3031   (aka M~81), would   have  $L_{bol}/L_{Edd} <
10^{-3}$.  Intriguingly, not only these two  sources lie in the region
of LINERs in the plane of Fig.  \ref{leddlr}, but  they are also close
to the separation  between Seyferts and LINERs  in the diagnostic line
ratios diagrams  \citep{ho97}.   NGC~3031 was  indeed considered as  a
LINER in the original classification of \citet{heckman80}.

On the   other  hand, for  both   LINERs and  FR~I nuclei  a  standard
accretion disk does not account for the low  fraction of the Eddington
luminosity observed  in the optical.   Thus, as already  discussed for
FR~I in CCC99,  our analysis  of the optical   nuclei shows that   low
radiative   efficiency accretion   processes  are  predominant  around
LINERs'  central  black holes.  This is  also  in agreement with other
independent evidence, such as the low X-ray luminosity of their nuclei
(e.g. Terashima et al. 2002).

More  importantly, the correlation between the  radio and optical core
of FR~I provides  a {\it minimum} to the  optical flux, represented by
the optical counterpart of the  non-thermal radio core.  Therefore, it
is  tempting  to interpret  the  ``optical excess''   observed in  the
``radio-quiet''   subclass of  LINERs  as optical  radiation from  the
accretion flow, in analogy with the  optical excess shown by Seyferts.
Thus, in  these  objects we  are probably  directly  observing optical
emission from a radiatively inefficient (ADAF-like) process.

\subsection{Radio-loudness, black hole mass and properties of accretion}

\begin{figure}[h]
\epsscale{1} \plotone{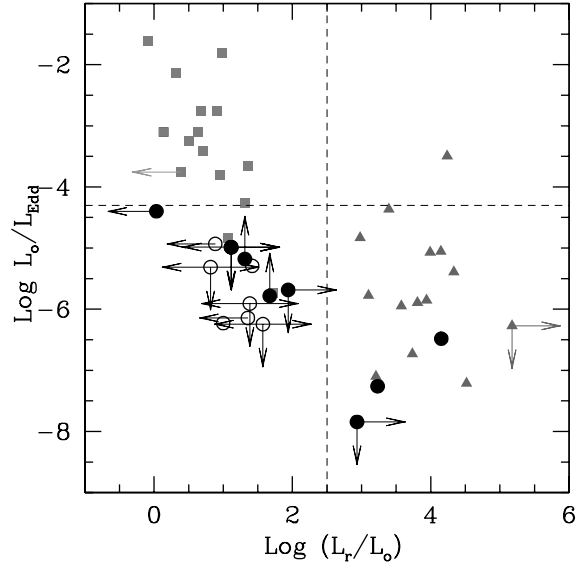}
\caption{$L_o/L_{Edd}$ vs. radio loudness of the nuclei for the 
three samples of local AGN. LINERs are marked as circles, Seyferts as  squares  and  FR~I  radio   galaxies
as triangles. Open  circles  are LINERs  in
late-type hosts, filled circles are LINERs in early-type hosts. Two LINERs (NGC~4450 and NGC~4550) overlap in this plot. The LINER NGC~1052 and the radio galaxy 3C~296 also overlap.}
\label{leddlrloud}
\end{figure}

As  discussed above,   LINERs    seem  to  appear   in   two  flavors:
``radio-quiet'' and ``radio-loud''.   The  two classes are similar  to
Seyferts   and FR~I radio  galaxies,  respectively, but scaled towards
lower  luminosities. For both  classes of LINERs the accretion process
is likely to be in the form of a low  radiative efficient disk and/or with a
low accretion rate. In this  section we investigate how the properties
of the accretion flow relate to  the radio-loudness parameter, for the
three samples of local AGN.

In Fig.  \ref{leddlrloud} we plot $L_o/L_{Edd}$ vs.  $R=L_r/L_o$.  The
sources  occupy different  regions of  the plane,  according  to their
physical properties.   We have divided the plane  into four quadrants,
which are not necessarily indicative of any physical state transition,
but    only    guide   the    eye    to    identify   the    different
classes. Nevertheless, the vertical  line, corresponding to $L_r/L_o =
2.5$ may  be interpreted as  our new definition for  the ``threshold''
between  radio-quiet  and  radio-loud  nuclei.  Seyferts  are  in  the
top--left part  of the diagram, which means  relatively high radiative
efficiency of the accretion  process (and/or high accretion rate), and
a low value for the radio-loudness parameter.

At least  seven LINERs (namely NGC~404,  NGC~2681, NGC~3368, NGC~3718,
NGC~4143 and NGC~4203, NGC~4736) lie on the radio-quiet side, and have
a low $L_o/L_{Edd}$ ($\lta  10^{-5}$).  The properties of these nuclei
are  independent of  the host  galaxy morphological  type,  since both
ellipticals  and spirals are  present in  this quadrant.   However, we
must point out  that the majority of ``radio-quiet''  LINERs have late
type  hosts.  Our  result show  that ``radio-quiet''  LINERs represent
Seyferts' low-efficiency counterparts.   As shown in Fig.\ref{bhvsrl},
these two classes also share the  same range of black hole masses.  In
the large  majority of the  models proposed to describe  the accretion
processes,  a  low  radiative  efficiency  is associated  with  a  low
accretion rate onto  the central black hole.  Thus  the most plausible
scenario is  that the  difference between Seyferts  and LINERs  of the
radio-quiet  type  resides in  the  accretion rate:
for a  given mass of the  central black hole, a  higher accretion rate
generates  a Seyfert nucleus,  while for  a low,  highly sub-Eddington
regime a LINER  is observed.  Furthermore it is  tempting to speculate
that the  properties of the accretion  regime may be  reflected in the
different   ionization   states    observed   in   the   two   classes
(low-ionization for  LINERs and high  ionization for Seyferts).  It is
worth noticing  that a small subsample  of LINERs might  be powered by
non-AGN processes  such as nuclear  star formation (e.g.  Maoz  et al.
1998). In particular,  this might be the case for  the nuclei that are
not detected in  the radio band and for which an  optical core is seen
(NGC~404 and  NGC~3368).  In these  objects, the optical  point should
then be interpreted as an upper limit to the AGN luminosity (i.e.  the
luminosity  of   the  accretion  process).   The   location  of  their
representative points would then be undetermined in the $L_r/L_o$ axis
(they   could  be   either  radio-loud   or  radio-quiet)   and  their
$L_o/L_{Edd}$  would be even  lower.  Therefore,  the scenario  is not
affected by the presence of a few non-AGN LINERs.

The lower-right region of the  plane is occupied by radio-loud nuclei,
which are either  LINERs or FR~I radio galaxies. At  least three of the
LINERs lie in this quadrant:  NGC~1052, NGC~4278 and NGC~4636 (which is
constrained  to  this region  by  the  non  detection of  its  optical
nucleus). By analogy with the FR~I, we identify the optical emission of
these nuclei as the optical  counterparts of the radio cores which, as
shown  by various  authors \citep{falcke,neil01},  are most
plausibly produced by non-thermal  synchrotron radiation. As for the
FR~I, the optical emission thus  represents an upper limit to the disk
component. 

Note that  although in Fig. \ref{leddlrloud} there   are many upper or
lower limits, almost all of them are in fact meaningful.  Even for the
three  LINER nuclei that  are  not detected  in  either the radio  and
optical bands, which results in a triple-direction limit, this implies
an upper limit for the  radiative efficiency of the accretion process.
Therefore,  although    we cannot establish   their    position on the
radio-loudness axis, their   location   on the $L/L_{Edd}$  axis   is
constrained to be significantly lower than the Seyferts, as observed
for all other LINERs of our sample.

\section{Conclusions}

We  have  explored  the   nature  of  LINERs  analyzing  archival  HST
observations and collecting radio data from the literature.  Of the 25
LINERs of our distance limited sample, nine show an unresolved nucleus
in the range  of wavelengths covered by the  HST observations. Four of
them  are detected  in the  optical, four  in the  near-UV and  one is
detected only  in the near-IR.   We have compared  their radio-optical
properties with those of the other unhidden AGN in the local universe,
i.e.  FR~I radio galaxies and Seyferts~1.

The radio-optical correlation found  for FR~I, which is best explained
as the  result of  a single  emission process in  the two  bands (i.e.
non-thermal  synchrotron emission  from the  base of  the relativistic
jet), provides  us with a powerful  tool to investigate  the origin of
the nuclei.

We have shown that in the  radio-optical plane of the nuclei there is
a clear  separation between Seyferts  and radio galaxies.   For similar
radio  core luminosity,  Seyfert~1 are  significantly brighter  in the
optical   than   FR~I.    Therefore,   although  most   Seyferts   have
$R=L_{5GHz}/L_B>10$, radio-quiet and radio-loud AGN appear to be still
well differentiated. This implies  that the nuclear physical properties of the
two classes are significantly different.

LINERs  appear to be   divided   into two subclasses, a   Seyfert-like
(radio-quiet)  group  and  a  radio  galaxy-like  (radio-loud)  group.
Although the number statistics are not sufficient to establish whether
there are two separate populations or there is a continuous transition
between the two, radio-quiet and radio-loud LINERs appear to represent
the  extension  of     Seyfert   and  radio-galaxies  towards    lower
luminosities, respectively.   This   is also mirrored by   their  host
galaxy  type.  Note that,   in this framework,  the ``radio-loudness''
parameter   acquires a physical   meaning, and indicates  what are the
dominant radiation mechanisms in the two bands.

The black  hole masses  of LINERs  span  the range $M_{BH}/M_{\odot} =
10^7-10^{8.5}$.  Although  we cannot derive  any correlation  with the
radio-loudness of the nucleus, it is apparent that radio galaxies host
the     more massive black  holes,  while   Seyferts'  black holes are
significantly less massive.  Intriguingly,  this separation holds also
between the two classes of  LINERS: ``radio-loud'' LINERs have central
black hole masses confined to $M_{BH}/M_{\odot} > 10^{8}$, while those
of ``radio quiet'' LINERs lie below such value.

We  have derived  the radiative  efficiency of  the  accretion process
around the  central black holes in  our samples of local  AGN.  All of
them emit only a small fraction of the Eddington luminosity.  Although
the determination of the bolometric luminosity is uncertain because of
the lack  of detailed spectral  information, the accretion  process in
LINERs  appears  to  take  place  on  a  highly  sub-Eddington  regime
($L_o/L_{Edd} < 10^{-5}$, and can  be as low as $\sim 10^{-8}$).  Such
low values  are clearly  not compatible with  the expectations  from a
standard   optically  thick   and  geometrically   thin  (quasar-like)
accretion  disk.   Thus, low  accretion  rates  and/or low  efficiency
processes  appears to  be required  in  all LINERs.   Our results  are
qualitatively  in agreement  with \citet{ho04},  who made  use  of the
$H\alpha$ emission line  as an indicator of the AGN  power for a large
sample of Seyferts and LINERs.

The  different  nature of the various  classes  of local  AGN are best
understood  when  the    fraction of the    Eddington luminosity  they
irradiate is plotted against the  nuclear radio-loudness parameter. In
this   diagram,   our objects  populate  three   different  quadrants,
according   to their physical  properties.  We   identify Seyferts and
radio-quiet LINERs as  the high and  low accretion  rate counterparts,
respectively.  For   low accretion regimes,  the  nuclei appears to be
``radio-loud'' only when  a more massive  black hole ($M_{BH} > 10^{8}
M_\odot$)  is  present. We speculate that  the  fourth quadrant, which
appears    to be  ``empty''  in   the   local universe, would  contain
radio-loud  nuclei  with high $L_o/L_{Edd}$,  readily  identified with
radio loud quasars.

The scenario we propose  needs  further investigation, since   optical
detections of the nuclei are  available for a  minority of the LINER's
sample.  Thus, deep imaging  with high spatial  resolution (achievable
only with the Hubble Space Telescope) are crucial. In particular, when
suitable observations of  a large number  of LINERs will be available,
it will be possible to address whether  the dichotomy persists or some
of the low-black hole mass objects with optical  upper limits mix with
the population of ``radio-loud''  LINERs.  Clearly this would  falsify
our scenario  for the role  of the black hole  mass in determining the
radio-loudness  of  the nuclei.   Deep  imaging  of a  larger complete
sample would  also address  the issue  of whether there  is continuous
transition between the two classes. This  is indeed a subject of great
interest in  the  study of high  luminosity  quasars  and for   a more
complete understanding of the  overall subject should be also extended
to lower end of the AGN luminosity function.

\begin{acknowledgements}

The authors wish to thank  Annalisa Celotti, Neil Nagar and William B.
Sparks  for insightful discussion.   We are  indebted to  the referee,
Yuichi  Terashima,  for  carefully  reading  the  manuscript  and  for
providing comments  and suggestions that greately  improved the paper.
MC  thanks  Gabriele   Ghisellini  for  insightful  discussions.  This
research has  made use of  the NASA/IPAC Extragalactic  Database (NED)
which  is  operated  by  the  Jet  Propulsion  Laboratory,  California
Institute of Technology, under  contract with the National Aeronautics
and  Space Administration.   MC  acknowledges support  from the  STScI
Visitor Program.

\end{acknowledgements}

\clearpage


\clearpage

\clearpage

\end{document}